\documentclass[12pt,nofootinbib]{revtex4}

\usepackage{amsmath,amsfonts,palatino,amsthm,epsf}

\newcommand{\beqa}{\begin{eqnarray}}
\newcommand{\eeqa}{\end{eqnarray}}
\newcommand{\f}{\begin{equation}}
\newcommand{\ff}{\end{equation}}

\newcommand{\bean}{\begin{eqnarray*}}
\newcommand{\eean}{\end{eqnarray*}}
\newcommand{\ra}{\rightarrow}

\usepackage{mathbbol}
\usepackage{amsmath}
\usepackage{amssymb}
\usepackage{dcolumn}
\usepackage{graphicx}
\usepackage{wasysym}

 \def\HH{{\cal H}}
\def\LL{{\cal L}}
 
\def\be{\begin{equation}} \def\ee{\end{equation}}

\begin{document}
\title{Quantum Graphity }

\author{Tomasz Konopka, Fotini Markopoulou and Lee Smolin}
\affiliation{Perimeter Institute, Waterloo, Canada, \\ University of
Waterloo, Waterloo, Canada}

\begin{abstract}
We introduce a new model of background independent physics in which
the degrees of freedom live on a complete graph and the physics is
invariant under the permutations of all the points. We argue that
the model has a low energy phase in which physics on a low
dimensional lattice emerges and the permutation symmetry is broken
to the translation group of that lattice. In the high temperature,
or disordered, phase the permutation symmetry is respected and the
average distance between degrees of freedom is small. This may serve
as a tractable model for the emergence of classical geometry in
background independent models of spacetime. We use this model to
argue for a cosmological scenario in which the universe underwent a
transition from the high to the low temperature phase, thus avoiding
the horizon problem. \end{abstract} \vfill

\maketitle

\tableofcontents

\newpage
\section{Introduction}

Background independent theories of quantum gravity are those that
share with general relativity the property that the formulation of
the laws of the theory does not require the specification of any
classical metric geometry or boundary conditions. As a result, such
theories are often formulated in the language of combinatorics and
representation theory, as that is what is left of quantum mechanics
when reference to manifolds and global symmetries is removed.

The biggest problem such approaches face is to demonstrate that
classical general relativity emerges from them at low energies or
large volumes.  There is some evidence for this, in several
approaches including causal dynamical triangulations \cite{CDT},
spin foam models \cite{SF} and loop quantum gravity \cite{LQG}, but
the problem is certainly not yet definitively solved.

One purpose of this paper is to suggest that it may help to take a
more physical approach to this problem. This begins by positing that
interesting models of quantum spacetime will have at least two
thermodynamic phases. In the high temperature, or disordered phase,
notions of geometry and perhaps even dimension and topology are
useless and the physics must be described in purely quantum
mechanical terms.  In the low temperature phase, the system becomes
ordered in such a way that it can be described in terms of fields
living on a low dimensional spacetime manifold with metric obeying
Einstein's equations, to a suitable approximation.

Seeing it this way may be helpful, because it may allow us to attack
the problem of the emergence of spacetime in the low temperature
phase with tools from statistical physics.  But there is another
reason for interest in such a scenario, which is that we know that
the universe has cooled from an initially very hot state. It is then
possible that the universe at some early time was in the high
temperature phase  and underwent a transition to the low temperature
phase at some time $t_c$.  Since geometry is an emergent property of
the low temperature phase we can call the transition {\it
``geometrogenesis."}  This event may have set up conditions which
are observable now in detailed observations of the cosmic microwave
background and large scale structure. If so it may be  possible that
such a scenario provides an alternative to inflation as an
explanation of the cosmological observations.

For example, in loop quantum gravity, the states are described by
graphs. Typical graphs, in this theory, as well as generally, have
high interconnectivity and do not admit of an easy description in
terms of a discrete geometry, nor do they easily embed in, or coarse
grain to, low dimensional geometries. They have small diameters (the
maximal distance between nodes, counted by graph links separating
them.) It is then natural to assume that the high temperature phase
is dominated by such non-geometrical highly connected graphs.  We
may note that this may play the role of an inflationary phase, in
ensuring that when the classical spacetime emerges, all regions of
space arise from parts of the graph that were in causal contact in
the earlier phase.

The physical question is then why graphs of low connectivity, low
valence and large diameter should dominate in the low temperature
phase. A second question is whether this scenario has consequences for cosmological observations.

Still another set of questions has to do with the role of emergent
symmetry in the phase transition to the ordered phase. In
\cite{Micromacro} it was shown that many models of dynamical quantum
geometry have emergent degrees of freedom which constitute noiseless
subsystems. These exist due to emergent symmetries which become
apparent only when the quantum system is analyzed by dividing it
into subsystems and environment.  These emergent particles carry
conserved quantum numbers and hence, as described in \cite{KM}, are
candidates for elementary particles. One proposal for how space
emerges is then that it is defined by the interactions of these
emergent particles. Using this language, we can then anticipate that
the transition to the low temperature phase will be accompanied by
an expansion in the Hilbert space dimensions of these noiseless
subsystems, corresponding to the emergence of translational and
rotational invariances.

Still another question raised by the scenario just discussed is
whether the transition to an ordered phase characterized  by the
emergence of local geometrical structure must be complete. Is it
possible that after the transition  there will remain defects in or
disorderings of locality\cite{Micromacro}?
What this means is that the state after
the phase transition may be dominated by graphs which only
approximately embed in low dimensional geometries. Defects in
locality would arise when two nodes of a low dimensional graph,
which are far away in the approximate classical metric, are
connected. One way to say this is that the notion of locality
encoded in the graph may not completely coincide with the notion of
locality given by the emergent metric that describes its course
grained properties.

In \cite{CFL} the implications of this possibility for cosmology are
investigated and it is found that there may indeed be striking
observational consequences of disordeded locality. This leads us to
ask another question about the phase transition from which space
emerges: is it possible that the result is a universe with
disordered locality?

To investigate all these questions, we decided to invent a model
which captures the key features of the scenario we have just
described, while being easier to analyze than full quantum gravity
models such as spin foam models.  The purpose of this paper is to
describe such a model and begin the investigation of its properties.

The model described here, which we call {\it ``quantum graphity,''}
is based on the complete graph on $N$ nodes. This means that every
two nodes in our graph are connected by an edge. The degrees of
freedom live on the edges of the graph and the dynamics is invariant
under the group of permutations of the $N$ nodes. There is a ground
state for each edge which signifies that the edge is turned off, and
excited states which indicate that the edge is on and in various
states. In the model we discuss here, there are three ``on" states
for each edge corresponding to the states of a spin-one system.
Thus, the states of the system include every graph on $N$ nodes.

We choose the Hamiltonian so that the ground state of the model
breaks the permutation symmetry by the formation of a low
dimensional lattice. We argue (but do not prove)
that under certain conditions the
spins in the system can arrange themselves in regular, lattice-like
patterns at low temperatures. When the graph is frozen, the model is
closely related to a model of Levin and Wen
\cite{Levin:2003ws,Levin:2004mi,Levin:qether} which has emergent
gauge degrees of freedom. The excitations of the spin system are
interpreted as photons coupled to massive charged particles and
propagating on the graph consisting of the ``on'' edges of the
graph.

The outline of the paper is as follows. In section \ref{s_model}, we
introduce a classical and a quantum model and explain the various
terms and constants in the proposed Hamiltonian. We discuss some
properties of the models in the high and low temperature regimes in
section \ref{s_phases}, and discuss the emergence of photons in
section \ref{s_u1}. The implications of the model for cosmology are
discussed in section \ref{s_cosmo}. We summarize in section
\ref{s_conclusion}.

\section{The model \label{s_model}}

In this section, we introduce the ``quantum graphity'' model based
on the complete graph of $N$ nodes. We begin by describing a
classical model and then extend it to a quantum mechanical setting.

\subsection{Classical model\label{s_classical}}

A complete graph on $N$ nodes is a collection of $N$ points, labeled
$a, b,\ldots$, which are all connected to each other by edges. We
make the edges carry labels $J$ and $M$ in the following possible
configurations \be (J,\, M) \in \{ (0,\,0),\, (1,\, -1),\,
(1,\,0),\, (1,\,1) \}. \ee We interpret the label $(0,\, 0)$ to
signify there is no link between two points, and the remaining
labels to signify there is a link.  We then call the state $(0, \
0)$ to be an ``off" state, while the three remaining states are
``on" states.

We consider a Hamiltonian
\be H = H_{links} + H_{vertices} + H_{loops} +
H_{hop}+H_{LQG}.
\ee
Let us explain the terms of $H$, in order. The
first term is
\be
H_{links} = V \, \sum_{a} \left(v_0- \sum_b
J_{ab}\right)^2;
\label{Hlinks}
\ee
where $V$ is a positive coupling constant,
$v_0$ is a fixed number, and the sums are over all points in the
graph. The minimum of $H_{links}$ occurs when the number of ``on''
links adjacent to every node in the graph is $v_0$. Thus this term
term tells us that the ground state will consist of graphs with
valence $v_0$.

The second term is
\be
\label{Hvertices} H_{vertices} = C
\sum_a \left( \sum_b M_{ab} \right)^2 + D \sum_{ab} M_{ab}^2
\ee
In the first line, the $C$ term favors configurations in
which the $m$ values of spins at each vertex add up to zero. The $D$
term gives preference to configurations in which all the said spins
have $m=0$.

The third term in the Hamiltonian is \be \label{Hl1} H_{loops} = -
\sum_{\mbox{minimal loops}} \frac{1}{L!} \, B(L) \prod_{i=1}^L M_i.
\ee Here, the sum is over minimal loops. We define these to be loops
that cannot be factored into the product of two loops of ``on" edges
that contain some of the same edges. The  products are defined over
a closed sequence of edges as follows \be \label{Mprod}
\prod_{i=1}^L M_i \, = \, M_{ab}\, M_{bc} \, \ldots M_{yz}\, M_{za}.
\ee $L$ is the length of the loop, and we note that a given on graph
defined  by an assignment of $J$'s can have mimimal loops of varying
lengths.

The coupling $B(L)$ is assumed to take the form \be \label{BLdef}
B(L) = B_0 B^L \ee where $B_0$ is a positive coupling constant and
$B$ is dimensionless. The separation of $B$ from $B_0$ is useful
because $B$ can be now associated with each instance of $M$ in the
loop product (\ref{Mprod}). We note then that the overall
coefficient of a loop term is proportional to $B^L/L!$. It is thus
small at very low and at very high $L$, but has a maximum value at
some particular $L_*$, \be \label{Lstardef} \frac{B^{L_*}}{L_*!} >
\frac{B^{L^\prime}}{L^\prime!} \qquad \forall L^\prime \neq L_*.\ee
We call $L_*$ the preferred loop length.

In comparison with $H_{links}$, note that $H_{loops}$ has an overall
minus sign. Note also that $H_{loops}$ contributes nothing to the
energy unless the edges it acts upon are in one of the ``on''
states. Thus there is a competition between this term and
$H_{links}$ which will be responsible for fixing the assignment of
``on'' edges.

\begin{figure}
\begin{center}
  \includegraphics[scale=1]{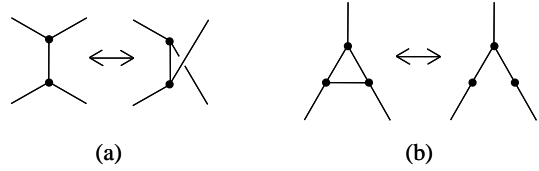}
\end{center}
\caption{ \label{f_moves} Examples of terms in Hamiltonian $\protect
H_{LQG}$ that acts on $\protect j$ variables. $\protect (a)$
Exchange of neighboring links. $\protect (b)$ Addition or
subtraction of an edge.}
\end{figure}

The last two terms in the Hamiltonian are $H_{hop}$ and $H_{LQG}$.
We will not need to specify these in detail, except to say that
$H_{hop}$ allows the $M$ variables to propagate from an on edge to
other edges adjacent to the same vertex, and the term $H_{LQG}$
generates local moves that turn edges on and off and thus allow one
configuration of ``on'' links to morph into another one. The action
of these graph-changing terms is illustrated in Figure
\ref{f_moves}. We assume that the couplings characterizing these
terms are such that they do not significantly alter the equilibrium
and ground states of the model.

\subsection{Quantum model\label{s_quantum}}

We now introduce a quantum model similar to the one we have just
described by turning the configuration space into a Hilbert space.
On each edge we put a four state Hilbert space, ${\cal H}_{spin}$,
which is spanned by an orthonormal basis of states $|j,\, m\rangle$,
\be \HH_{spin} = \mathrm{span} \, \{ |0, 0\rangle, \, |1, -1\rangle,
\, |1, 0\rangle, \, |1, +1\rangle \}. \ee Since there are $N(N-1)/2$
links in a complete graph with $N$ points, the Hilbert space for the
whole system is, \be \HH_{tot} = \otimes^{N(N-1)/2} \, \HH_{spin}.
\ee As in the classical model, we will interpret states with
$j_{ab}=0$ ($j_{ab}=1$) as indicating the absence (presence) of a
link between points $a$ and $b$.

We define four operators acting on the Hilbert space of each
spin\footnote{The operators we define are related to the angular
momentum operators $J^2$, $J_z$, and $J^\pm$ in their usual
defintions. The two sets of operators differ however in their
normalization by factors of $\sqrt{2}$, which in the present setup
is equal to $\sqrt{j_{max}(j_{max}+1)}$.}. The first two, $J$ and
$M$, are eigen-operators of the states $|j, \, m\rangle$ such that
\be
\begin{split} J\, |j, \, m\rangle &= j\, |j,\, m\rangle \\
M\, |j, \, m\rangle &= m\,|j,\, m\rangle.
\end{split} \ee The other two are the lowering operator $M^-$ and
the raising operator $M^+$ which act as \be \begin{split}
\sqrt{2}\, M^+ \, |j, \, m\rangle &= \sqrt{(j-m)(j+m+1)} \, |j, \, m+1\rangle \\
\sqrt{2}\, M^- \, |j, \, m\rangle &= \sqrt{(j+m)(j-m+1)} \, |j, \,
m-1\rangle.
\end{split} \ee
In terms of algebra, the operator $J$ commutes with $M$ and $M^\pm$,
which form a closed algebra among themselves \be [M^+, \,M^-] = M,
\qquad [M,\,M^{\pm}] = \pm M^{\pm}. \ee It will be important later
that all these operators annihilate the $|0,\, 0\rangle$ state, \be
J\, |0,\,0\rangle = M\,|0,\, 0\rangle = M^\pm \, |0,\, 0\rangle = 0,
\ee and that all non-zero matrix elements have unit magnitude.

We now write the quantum Hamiltonian
\be
 \hat{H} = \hat{H}_{links} + \hat{H}_{vertices} +\hat{H}_{loops} +
\hat{H}_{hop} + \hat{H}_{LQG}. \ee The first terms $\hat{H}_{links}
+ \hat{H}_{vertices}$  are gotten from the classical (\ref{Hlinks})
and (\ref{Hvertices}) by a replacement of classical values with
quantum operators.

The next term, the quantum loop Hamiltonian, $\hat{H}_{loops}$, is
\be \label{Hloops} \hat{H}_{loops} = - \sum_{\mbox{loops}}
\frac{1}{L!} \, B(L) \prod_{i=1}^L M^{\pm}_i \ee now involves sums
over loops of varying even lengths, $L$.  It is slightly different
than its classical counterpart in that the product of $M^{\pm}_i$ is
understood as \be \label{Mseries} \prod_{i=1}^L M^\pm_i \, = \,
M^+_{ab}M^-_{bc} \ldots M^+_{yz}M^-_{za}. \ee As before, the
products of operators $M^\pm$ act on successive links along minimal
links of a loop. Since the series (\ref{Mseries}) starts with $M^+$
and ends with $M^-$, the length $L$ of the loop $a, b, \ldots, z$
must now be even, $ L= 4,\, 6,\,8, \, \ldots$. We  may simplify the
model by restricting to low $L$, for example, $L=4,6$.
 We chose the coupling $B(L)$ to be of the same
form as in (\ref{BLdef}) and define the preferred loop length $L_*$
analogously to (\ref{Lstardef}).

We do not specify the details of the final two terms in $\hat{H}.$
Their role is again to move $M$ values between neighboring edges
($\hat{H}_{hop}$) and allow the graph to morph from one
configuration to another ($\hat{H}_{LQG}$) by moves shown in Figure
\ref{f_moves}.

This quantum model is significantly more complex than the classical
one in the previous section. One of the complications is that the
loop Hamiltonian $\hat{H}_{loops}$ does not commute with
$\hat{H}_{vertices}$, which implies that eigenstates of the
Hamiltonian will generally be superpositions of states involving
different $m$ configurations. To understand the role of the terms
$\hat{H}_{vertices}$ and $\hat{H}_{loops}$, it is helpful to first
consider the graph of ``on'' links to be frozen in a particular
configuration, say a regular cubic lattice where the minimal loops
in the graph are plaquettes. In this case, these terms reduce to the
rotor model of Levin and Wen \cite{Levin:qether}. We briefly
describe the expected physics as this reduction will become
important in section \ref{s_u1} where we will discuss the $m$
degrees of freedom as giving rise to a gauge theory on a lattice.

In the absence of the loop term, the ground state of
$\hat{H}_{vertices}$ consists of all links having $m=0$. Excited
states appear as open or closed chains of alternating $m=+1$ and
$m=-1$ links. These excitations are called strings. Their energy
above the ground state is proportional to $D$ times the number of
edges forming the string. Thus the coupling $D$ can be thought of as
a string tension. Nodes on which the $C$ term is not minimized are
said to carry the ends of open strings. The energy of a string end
is proportional to $C$. We take $C \gg D$ such that open ends occur
very infrequently
 or not at all.

Given a graph with all ``on'' edges labelled by $m=0$, a loop
operator acts as to create a closed string of alternating $m=+1$ and
$m=-1$ links. The string will acquire tension through the $D$ term.
However, since the sign of the $B_0$ term is negative, the overall
energy of the state may increase or decrease and this creates the
possibility of two distinct scenarios. In one scenario, the tension
in a string is greater than the contribution from the loops term, so
the overall effect of creating a string is to increase the energy of
the system. If this is the case, then the string represents an
excited state over the vacuum in which all $m$ values set to zero.
The second scenario is the one that we will be mostly interested in.
There, the tension is small compared to the contribution from
$\hat{H}_{loops}$ so that creating a string decreases the energy.
This indicates that the true ground state of the model consists of a
superposition of a large number of strings, a string condensate. We
should note that because the graph has a finite number of nodes and
the $m$ values on each edge only take three possible values, the
Hamiltonian is bounded from below. Hence the string-condensed ground
state exists even though it is difficult to write down.

The quantum model described in this section is not defined on a
regular and fixed lattice - all the terms in the Hamiltonian are
invariant under permutations. Nevertheless, we can expect the same
kind of competition between the $D$ term and the $B_0$ term to
possibly lead to string condensation.

\medskip

To summarize, our proposed quantum model contains four dimensionful
coupling constants ($V, C, D,$ and $B_0$) and three dimensionless
numbers ($N,$ $B$ and $v_0$). These parameters are presumed to
satisfy some rather generic conditions such as $N \gg 1$, $C \gg D,$
and $V\gg 1$. We discuss these conditions in more detail in the next
section.

\section{Thermodynamic Phases \label{s_phases}}

We now describe the states of the graph when it is coupled to a heat
bath with temperature $kT=1/\beta$. As the model contains several
coupling constants, the complete phase diagram of the model is
expected to be very complex. We focus only on two extreme regimes -
the very high and the very low temperature regimes - and, although
we do not study their nature, suppose there to be one or more phase
transitions that occur between them.  We will assume for simplity
also that $V >> B,C,D$ so that the dominant term at high temperature
comes from $H_{links}$.

\subsection{High Temperature}

When the temperature is high, $T>V$, the spins are in a disordered
state and the average valence of a typical member of the thermal
ensemble can be high. In this high temperature regime the system can
be thought of as non-local as thermal fluctuations are capable to
turning on or off links between any two points.

When $T \gg V$, we can ignore the terms in the Hamiltonian that
depend on $M$. Since the spins fluctuate, we can also approximate
the partition function $Z_N$ for the whole system by $H_{link}$ alone, which means that we may take
\be
\label{ZNapprox} Z_N \approx Z_1^N,
\ee
where $Z_1$ is the parition
function for one node only. We write $Z_1$ as \be Z_1 = \sum_v 3^v
\, \mathrm{exp}\left(-\beta V (v-v_0)^2\right) \ee with the sum over
all possible valences. Since each ``on'' spin can take three
possible $m$ values, there is a multiplicity factor $3^v$ in front
of the Boltzman factor. We further write \be Z_1 = \sum_v
\mathrm{exp} \left( -\beta f(v) \right) \ee as a definition of
$f(v)$. It follows that \be f(v) = V(v-v_0)^2- vkT \, \ln 3 \ee and
that its minimum occurs when \be \label{veffective} v = v_0 +
\frac{kT}{2V} \ln 3. \ee Thus we find that the effective valence of
a point increases linearly with temperature.

A typical graph in the ensemble then may be considered to be a
random graph, characterized by a probability $p$ that each of the
$N(N-1)/2$ of the edges are excited. Given (\ref{veffective}) and
the fact that there are $N-1$ edges adjacent to each node, this
probability is thus \be p = \frac{v}{(N-1)} \sim \frac{T}{NV}. \ee
In random graph theory there is a probability $p_0 \sim 1/N$ that
marks the transition above which a typical graph is connected. The
ensemble will be in this regime so long as $T > T_{connected} \sim
V$. When the temperature $T$ is on the order of $NV$, almost all
edges in the complete graph will be ``on.''


\subsection{Low Temperature}

When the temperature is very low, $T \ll V$, the graph will
essentially consist of nodes with a fixed valence $v_0$. Once the
links Hamiltonian is at its minimum, the other terms
$\hat{H}_{vertices}$ and $\hat{H}_{loops}$ gain importance. We now
proceed to discuss the effect of those $M$ sensitive terms in the
Hamiltonian on the distribution of ``on'' links. At the same time,
we assume that the amplitudes of the LQG terms are very small so
that any low lying excitations do not change the patterns of ``on''
and ``off'' links. Instead the low lying excitations involve
propagation of the $m$ degrees of freedom. Hence for the rest of
this section, we use the terms graph and network to refer to
configurations of ``on'' links. We focus on the case $v_0=4$ for
concreteness.

\begin{figure}
\begin{center}
  \includegraphics[scale=1]{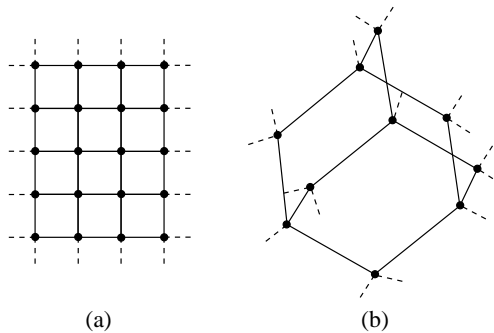}
\end{center}
\caption{ \label{f_3dshapes} Sample lattice formations with 4-valent
vertices. Solid lines represent ``on'' links and dashed lines
designate how the fragments shown fit inside a larger lattice.
``Off'' links are not shown.}
\end{figure}

There are many distinct types of structures that can be built out of
a 4-valent network of edges. There can be tree-like structures in
which there are no closed loops, regular structures made out of
polygons with four, six, or more sides, or regular and irregular
structures made up of many different kinds of polygons. Graphs can
be planar or non-planar. Two of these graphs, the two-dimensional
square lattice and the three-dimensional diamond lattice, are shown
in Figure \ref{f_3dshapes} but other arrangements are possible as
well. Based on valence alone, all these structures are in principle
candidates for the ground state of the system at low temperatures.

The terms in $\hat{H}_{vertices}$ and $\hat{H}_{loops}$ can
differentiate between these candidates. In particular, loop terms
can lower the energy when the graph contains closed paths. In the
classical model, this can happen when the couplings $C$ and $D$ are
negligible and the $m$ values on the edges are set to $+1$ or $-1$,
or combinations of these. For a particular value of $B$, there will
be a preferred loop length $L_*$  and the ground state will be such
that the number of loops of this length is maximized. Given a
limited valence of the nodes, this maximization criterion may have
the effect of creating a low-dimensional lattice. For example, if we
consider only even-length loop terms and chose $B$ such that $L_* =
4$, the low dimensional graph may resemble the square lattice shown
in Figure \ref{f_3dshapes}(a). If we chose $B$ such that $L_* = 6$,
the diamond shaped lattice in Figure \ref{f_3dshapes}(b) would be
preferred.

In the quantum model, the loop terms generate strings of alternating
$m=+1$ and $m=-1$ edges which contribute a positive energy through
the tension $D$ term. For loops to be encouraged in the low
temperature limit, we need to have the loop term be dominant over
the tension. We can roughly work out when string condensation can
happen given a particular graph. As a first example, we take the
square grid shown in Figure \ref{f_3dshapes}. To consider this edge
configuration as a plausible configuration, we need to take $B$ such
that $L_* = 4$. On this graph there are two edges per plaquette.
Thus string condensation could occur only if the couplings satisfy
the inequality \be 2D - \frac{1}{4!}B_0 B^4 <0. \ee Alternatively,
we can introduce a parameter $\gamma
> 1$ and set \be \label{BDconditions} 48 \gamma D = B_0 B^4. \ee
When these requirements are satisfied, we can argue that the ground
state lattice should be square and that it should support string
condensation. As a different example, consider the diamond lattice
shown in Figure \ref{f_3dshapes}(b). Diamond is characterized by
hexagonal plaquettes, so we chose $B$ such that $L_* = 6$. The
requirement for string condensation should thus be \be 2D -
\frac{1}{6!}B_0 B^6 < 0 \ee because there are again two edges per
plaquette.

It is reasonable to ask whether the lattices considered in Figure
\ref{f_3dshapes} are truly the ground states of the model for the
parameter ranges specified. We cannot fully answer this question at
this point but we can offer additional observations which support
our proposal. If we think of the ground state lattice to crystallize
by evolving via the moves of Figure \ref{f_moves} and note that
these moves do not disconnect graphs and act equally on all possible
edge configurations, then we would be led to hypothesize that the
ground state lattice should be connected and homogenous. The
configurations in Figure \ref{f_3dshapes} thus seem to be good
candidates. The diamond lattice seems like a particularly good
candidate as each pair of edges at a vertex supports a plaquette.

We emphasize that, in the quantum model, the formation of a lattice
at low temperatures must be accompanied by string condensation. We
discuss what this means in more detail in the next section.

\section{The emergence of gauge fields at low temperature \label{s_u1}}

Once the link degrees of freedom freeze, we are left just with $m$
degrees of freedom. It turns out these give a lattice gauge theory
\cite{Levin:2003ws,Levin:2004mi,Levin:qether}.

We consider a phase in which the links are arranged in a regular
three-dimensional pattern of four-sided plaquettes and there are no
open ended strings. This could happen for example when the valence
of a node is set to $v_0 = 6$. In this phase, the $V$ and $C$ terms
in the Hamiltonian are constant and we ignore them\footnote{The term
proportional to $C$ can be understood as a mass term for a scalar
particle corresponding to an end of an open strings, see
\cite{Levin:qether} for details.}. The non-vanishing terms then
consist of the tension and loop operators with exactly four edges,
\be \label{Hleading} H_{lowT} \sim D \sum_{ab} M_{ab}^2 -
\frac{1}{4!}B_0 B^4 \sum_{a} \prod_{i=1} M^{\pm}_i. \ee Since the
lattice is regular, the sum over loops can be thought of as a sum
over plaquetes. We can define a plaquete operator $W_{a^\prime}$
anchored at a point $a^\prime$ as \be W_{a^\prime} = M^+_{a^\prime
b}M^-_{bc} M^+_{cd} M^-_{da^\prime}. \ee The points $a^\prime,
\ldots d$ are now fixed by a convention of labeling plaquettes given
their base point so that there is no summation over repeated
indices. With the help of this operator, the Hamiltonian can be
written as follows \be H_{low T} \sim D \sum_{ab} M_{ab}^2 -
\frac{1}{3!} B_0 B^4 \sum_{a^\prime} \left( W_{a^\prime} +
h.c.\right) \ee where $h.c.$ stands for the Hermitian conjugate of
$W_{a^\prime}$, i.e. a loop operator with $M^+$ and $M^-$
interchanged on each link. The sum in the second term is over
plaquettes.

It turns out that this Hamiltonian correspond to $U(1)$ gauge theory
in axial $A_0=0$ gauge. In fact, the Kogut-Susskind Hamiltonian
\cite{Kogut} for a gauge field on a cubic lattice in three spatial
dimensions is \be H_{KS} = \frac{g^2}{2a} {\sum_{ab}}^\prime
M_{ab}^2 - \frac{2}{ag^2} \sum_{a^\prime} (W_{a^\prime} + h.c.). \ee
The sum in the first term is shown primed because it is only over
nearest neighbors connection in the lattice. The variables $a$ and
$g$ denote the lattice spacing and the coupling constant,
respectively. Comparing coefficients of our model and the gauge
theory Hamiltonian gives the identifications \be \label{iden} g^2
\sim \sqrt{\frac{4!D}{B_0 B^4}}, \qquad \frac{1}{a^{2}} \sim
\frac{1}{3!}D B_0 B^4. \ee

Recall that string-net condensation, and thus the possibility of
emergent $U(1)$ theory, only happens when certain conditions such as
(\ref{BDconditions}) are satisfied. Inserting (\ref{BDconditions})
into the expression (\ref{iden}) for $g^2$ above gives \be g^2 =
(2\gamma)^{-1/2}. \ee Recall that $\gamma > 1$, so that the coupling
$g^2$ of the emergent gauge field is weak. Furthermore, if we set
the inverse lattice spacing on the order of the Planck mass, $
a^{-1} \sim m_P, $ then it follows from (\ref{iden}) and
(\ref{BDconditions}) that \be \label{Bfound} B_0 B^4 \sim m_P
g^{-2}, \qquad D \sim \, g^2 \, m_P. \ee

To be sure, the correspondence between $H_{lowT}$ and $H_{KS}$ is
not exact because in the pure gauge theory, the edges connecting
lattice sites can carry arbitrary representations of $U(1)$ whereas
the $m$ labels in our model can only take three values $-1, 0,$ and
$1$. Nonetheless, the string condensed phase of $H_{lowT}$ should
exhibit features of $U(1)$ theory, such as the presence of
photon-like excitations, even in this crude approximation
\cite{Wenprivate}. The correspondence could be improved by allowing
a wider range of $m$ values on each edge. This is not in principle
an obstacle for our model, but we keep the present setup for
simplicity.

\section{Quantum graphity as a model of the very early universe \label{s_cosmo}}

Based on the quantum graphity model we have described in this paper
we can propose the following scenario for the early history of the
universe.

At early times, when $T \gg V$, the graph is in a very disordered
state and the average valence of each node is large. The diameter of
the graph, or the distance measured in ``on'' links between any two
points, is approximately $\log (N)$ in this phase so that the
degrees of freedom on a typical graph quickly come into thermal
contact. Thus, the whole system may be assumed to come into thermal
equilibrium.

As the system cools and the temperature drops, however, one or more
phase transitions may occur in which the $j$ degrees of freedom will
become frozen. How the system cools depends on the relations between
different coupling constants. We assume that the first transition
that occurs is one in which the valence of each node becomes frozen
to $v_0$. Thus as the temperature cools below $V$ the spins arrange
themselves into regular patterns that can be interpreted as extended
space. We thus have the emergence of classical geometry as well as
standard gauge theory and matter fields.

Even if this is a simplified model of the emergence of space, it
suggests insights for physical cosmology. The
horizon problem is the statement that distant parts of the universe
appear to be in thermal equilibrium despite universe's evolution
suggesting that these parts could not have interacted during the
course of the universe's estimated lifetime. This is deemed to be a
puzzle because its most straight-forward resolution by positing
special initial conditions lacks physical justification. Our model
provides such a justification because it suggests that the spins
were part of a thermal ensemble before the temperature fell
sufficiently for the system to enter a phase of classical geometry.
Thus the model shows the horizon problem may be avoided if geometry
is emergent. In this sense, the model also provides an explicit
example of a broader idea that a distinction between micro-locality
(locality between fundamental degrees of freedom) and macro-locality
(locality between emergent degrees of freedom) may be important for
understanding quantum gravity and the physics of the very early
universe \cite{Micromacro}.

The model also allows us to discuss an important issue for quantum
gravity models, which is the role of diffeomorphism invariance. When
a notion of geometry emerges in the low energy limit, each state of
the system corresponds to a geometry, which is a description of
metric and fields modulo diffeomorphisms. This is because there is
no role for diffeomorphism transformations in the original model,
because it makes no reference to geometry.

\begin{figure}
\begin{center}
  \includegraphics[scale=1]{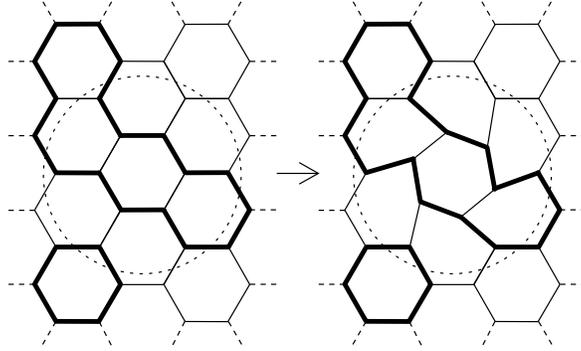}
\end{center}
\caption{ \label{f_diffeos} Diffeomorphism invariance on the
honeycomb lattice. The map is trivial outside the dotted region and
is a twirl or rotation inside that region. Bold lines indicate
strings.}
\end{figure}

To see this in detail we consider the case of the honeycomb lattice.
We notice it can be naturally embedded in a flat two-dimensional
space. Thus, it is possible to associate to it a flat Minkowski
metric $g$ which assigns lengths to the spins and positions to the
points. (In fact, we made this assignment in the previous analysis
by defining a spacing variable $a$.) Consider now a diffeomorphism
$\varphi$ acting on $g$. The invariance of the spin model under such
deformations is illustrated in a concrete example of a $\varphi$
with compact support in Fig. \ref{f_diffeos}, where we see that
$\varphi$ moves the links and vertices but preserves the
connectivity of the various elements. Moreover, the relation between
excitations is unchanged, and since observables can only be defined
in terms of relative excitations and connectivity of the graph
without reference to the embedding, the dynamics and predictions
made before or after the diffeomorphism are the same.

\section{Conclusions \label{s_conclusion}}

In this paper we have presented a class of models in which there is
no notion of geometry fundamentally, but which allow geometry to
emerge as an approximate description of a low temperature phase.
We have conjectured that the low temperature phase is characterized by the formation of large, regular lattices. While we motivated this conjecture more work would need to be done to test it.

It is interesting to compare our model to the usual theory of
gravity and matter defined by the Einstein-Hilbert and matter
Lagrangian, \be \label{S_Hilbert_Matter} S = \frac{1}{4\pi G} \int
d^4x \sqrt{-g} \left( R + 2\Lambda\right) + \int d^4x \sqrt{-g}
\,\LL_{M}. \ee This formalism admits several limits which enable to
study subparts of the full theory. On one hand, when the matter
component is ignored by setting $\LL_M = 0$, the leftover
Einstein-Hilbert action is well defined and describes the dynamics
of the gravitational field by itself. Indeed, there are many
interesting and physically relevant solutions to the pure
Einstein-Hilbert theory when $\LL_M=0$. On the other hand, the
no-gravity limit $G\ra 0$ is also well defined and describes a
quantum field theory on a fixed background $g_{\mu\nu}$.

This kind of splitting does not occur in our model because the
matter degrees of freedom play an essential role in organizing the
links into a regular lattice structure. We find that if we try to
construct a discrete space in a background independent manner, it is
helpful to assign both $j$ and $m$ variables to the spins and write
a Hamiltonian that acts on them individually. It turns out that the
dynamics of the $m$ variables that organizes the ``on'' links
automatically gives us a $U(1)$ gauge theory. Alternatively, if we
try to formulate a string condensation model of a $U(1)$ gauge
theory so that it does not depend on a fixed lattice, we find that
it is merely necessary to introduce one more state ($|0,\,
0\rangle$) in the Hilbert space of each spin and to fix the valence
of the nodes. Viewed in this way, the Hamiltonian of the quantum
model is rather economical.

Apart from constructing a geometry, we have not attempted to
reproduce features of gravitational physics such as gravitons in our
model. This could be attempted using again string condensation
techniques, following \cite{Gu:2006vw,Lee:2006gp}, or perhaps using
loop quantum gravity as a guide.

In the future we hope to use this model to study features of the
conjectured geometrogenesis phase transition.  As argued in
\cite{holocosmo} and \cite{CFL} some aspects of the transition
may be measurable. These include the transition temperature
and a critical exponent that governs the speed of the transition and the proportion of non-local links left over after the transition.

\begin{acknowledgments}
We would like to thank Ian Affleck, Olaf Dreyer, Stefan Hofmann,
Michael Levin, Joao Magueijo, Seth Major, Chanda Prescod-Weinstein, Xiao-Gang Wen and Hans
Westman for comments and conversations during the course of this
work.
\end{acknowledgments}

\end{document}